\documentclass[11pt]{article}

\usepackage[english]{babel}
\usepackage[letterpaper,top=1in,bottom=1in,left=1in,right=1in,marginparwidth=0.75in]{geometry}

\usepackage[version=4]{mhchem} 
\usepackage{siunitx}
\usepackage{graphicx}
\usepackage{dcolumn}
\usepackage{bm}
\usepackage{setspace}
\usepackage{amsmath}
\usepackage{fixmath}
\usepackage{graphicx}
\usepackage{amssymb}
\usepackage{mathtools}
\usepackage{cite}

\title{Optimizing Coded-Apertures for Depth-Resolved Diffraction}
\author{Do\u ga G\" ursoy\footnote{E-mail: dgursoy@anl.gov}, Dina Sheyfer, Michael Wojcik, Wenjun Liu, Jonathan Tischler \\
Advanced Photon Source, Argonne National Laboratory, \\
9700 S Cass Ave, Lemont, IL 60439, USA}

\begin{document}
\doublespacing
\maketitle

\begin{abstract}
Coded apertures, traditionally employed in x-ray astronomy for imaging celestial objects, are now being adapted for micro-scale applications, particularly in studying microscopic specimens with synchrotron light diffraction. In this paper, we focus on micro-coded aperture imaging and its capacity to accomplish depth-resolved micro-diffraction analysis within crystalline specimens. We study aperture specifications and scanning parameters by assessing characteristics like size, thickness, and patterns. Numerical experiments assist in assessing their impact on reconstruction quality. Empirical data from a Laue diffraction microscope at a synchrotron undulator beamline supports our findings. Overall, our results offer key insights for optimizing aperture design in advancing micro-scale diffraction imaging at synchrotrons. This study contributes insights to this expanding field and suggests significant advancements, especially when coupled with the enhanced flux anticipated from the global upgrades of synchrotron sources.
\end{abstract}

\section{Introduction}

The diffraction of a tightly focused light beam from a crystalline specimen has been instrumental in establishing the fundamentals of diffraction microscopy. Particularly, utilizing hard x-ray energies typically ranging from about ten to several tens of kiloelectronvolts enables the non-invasive imaging of three-dimensional (3D) volumes within bulk matter at micrometer length scales, thanks to the improved penetration depth and reduced scattering associated with these energies. These types of experiments are readily accessible at synchrotron facilities, many of which are currently being upgraded to offer brighter sources \cite{shin2021new}. 

Based on this principle, numerous microscopy techniques have emerged, each with distinct setups and illumination types. Notable techniques include Laue diffraction \cite{ice20003d, kunz2009dedicated, tamura2002submicron, ulrich2011new, hofmann2009probing, maabeta2006defect, park2007local}, x-ray scattering \cite{poulsen2004three, bernier2020high, tsai2021grazing}, and dark-field x-ray microscopy \cite{simons2015dark, jakobsen2019mapping, kutsal2019esrf}. These techniques are evolving alongside advancements in synchrotron science, leveraging its capabilities to provide valuable diffraction information. For example, Laue diffraction offer detailed insights into crystal lattice structures and defects, x-ray scattering provide structural understanding at molecular and nanoscale levels, and direct imaging methods such as dark-field x-ray microscopy enable direct visualizations of diffraction from a volume within bulk matter. Together with other methods, they enable our understanding of  mesoscale phenomena of ordered materials.

The data collection process of diffraction microscopes is similar and involves scanning the sample in the focused beam across a predefined grid. The objective is to collect diffraction data at each scanning point, thereby constructing a two-dimensional image of the sample assuming the sample is scanned both vertical and horizontally, mapping its diffraction properties into a two-dimensional (2D) image. Typically, the diffracted beams at each scanning point are captured using a pixel array area detector, facilitating analysis to yield insights into the structure of the crystal lattice, strain patterns, and orientations within the material along the path of illumination. 

To obtain a 3D image, one can either rotate the sample while raster scanning it and utilize tomographic reconstruction methods, or scan an aperture across diffracted beams. These approaches enable spatially resolved diffraction along the illumination path, a capability not achievable only through raster scanning the sample with a focused beam. Although both methods offer different strengths, aperture scanning is generally easier to implement and mechanically more robust, as it requires no or less movement of the sample compared to rotational scanning. Nevertheless, additional factors such as sample size, grains and deformations under study warrant careful consideration, often rendering certain methods more suitable than others.

This paper focuses on investigating apertures optimized for depth resolution, thus concentrating on aperture scanning methods. These apertures, which include highly absorbing wires, pinholes, narrow slits, or knife-edges, individually resolve the depth of diffracted beams \cite{Larson:02}. Functioning akin to a scanning pinhole camera, they selectively permit specific rays to pass through or be partly blocked by the aperture. While these types of apertures are commonly utilized, they often result in prolonged data acquisition times due to the necessity of scanning the aperture across the entire targeted sample region to capture all diffracted beams. For instance, certain experiments conducted at the Laue-diffraction microscope situated at the Advanced Photon Source (APS) may require hours, or even days, to collect sufficient data to understand a phenomenon \cite{song2023dose}.

\begin{figure}
\centering
\includegraphics{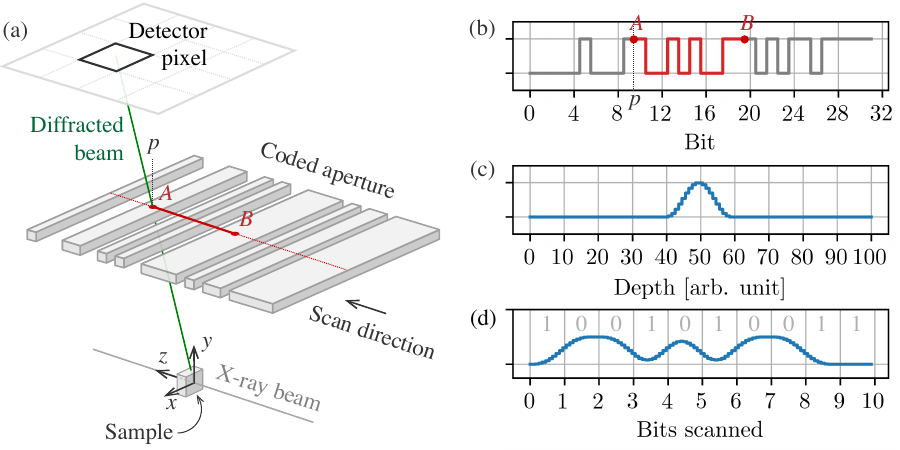}
\caption{(a) Setup for measuring diffracted beam via coded aperture scanning. (b) Coded aperture pattern modulating from point A to B. The position of the initial coded aperture is marked with $p$ (c) Signal from the sample along the x-ray beam. (d) Modulated intensity recorded in the detector pixel.}
\label{fig1}
\end{figure}

Recently, we introduced the concept of utilizing coded apertures and associated signal reconstruction methods as a more efficient alternative to conventional pinhole camera-like approaches, aimed at saving time without compromising precision \cite{Gursoy2022coded}. Coded apertures, specifically designed for micro-diffraction applications, function as light-absorbing elements with patterned structures that modulate incident light during scanning. The modulation is solely absorption-based and does not consider phase modulations of the beam. These patterns, typically pseudo-random in nature, require the aperture material to be partly opaque within the relevant energy range to effectively modulate the incident intensity on the detector. Current micro-fabrication techniques based on lithography allows to make solid apertures at desired length scales.

Figure~\ref{fig1} demonstrates the basics of operation principle. A specimen is illuminated by a focused x-ray beam, inducing diffraction that satisfies the Bragg angle. Scanning of the coded aperture across the diffracted beam results in intensity modulation in the detector pixel. In other words, the recorded intensity reflects the pattern of the aperture. Therefore, a sufficiently long scan reveals the direction of the beam by aligning the modulation pattern with that of the aperture. Since intensity modulation follows a linear process, resolving the shape of the beam requires solving a deconvolution problem, where the modulation pattern serves as the convolution kernel. This can be achieved through direct inversion or iterative methods such as the least squares technique. Such an approach facilitates spatially resolving the beam along its illumination path, facilitating depth-resolved diffraction. This process can be applied to each pixel on the array detector to reveal the corresponding signal emanating from the sample, thus forming a comprehensive understanding of the diffraction properties in 3D.

Coded apertures hold promise in diffraction microscopy, but their performance can vary significantly. One key factor lies in the patterning, which defines the encoding and decoding schemes. Other factors affecting the performance include bit sizes, incident angle, aperture thickness, and measurement noise, with the latter being particularly significant when using short pixel exposure times. In this paper, our goal is to assess the influences of these parameters on method's performance. To achieve this, we systematically studied aperture parameters through sensitivity analysis in numerical simulations and with real experimental data. These studies identified critical parameters for depth resolution and informed optimal experimental designs. To validate our findings, we conducted experiments using a Laue diffraction microscope on a synchrotron undulator beamline at the APS and compared our results with numerical simulations.

\section{Methods}

In this section, we outline the methodologies employed for simulating measurement data (forward model) and spatially reconstructing diffraction beams (inverse model). Furthermore, we illustrate the design parameters for coded-aperture and scanning utilized in numerical experiments, along with methods for quantifying these designs.

\subsection{Data simulation model}

We commence by introducing a parameterized model for simulating measurements, outlining the model parameters under investigation. Our study is centered on binary coded apertures, resembling barcode structures with linear bars representing bits. This choice is driven by the ease with which our current experimental setup can accommodate them. For a one-dimensional (1D) aperture function denoted as $a(z)$ along the scan axis, the intensity ($I$) of a diffracted beam recorded in a detector pixel can be modeled as follows
\begin{equation}
\label{eq1}
    I = \int_{\Omega} a(p - z) s(z) dz,
\end{equation}
where $p$ is the initial position of the intersection of the diffracted beam with the coded-aperture as shown in Figure~\ref{fig1} and marks the beginning of the coded-aperture segment for signal modulation, $s(z)$ represents a one-dimensional diffracted signal incident upon the coded aperture plane (or line for a 1D aperture), and $\Omega$ signifies the interval within which the signal exhibits non-zero values (e.g., $\Omega\in[A, B]$ for the illustration in Figure~\ref{fig1}). 

To resolve the direction of the signal, determining the positional parameter $p$ is required, while resolving the signal's shape, that is $s(z)$ or simply $s$, necessitates solving the deconvolution problem for a given $p$. The first problem of resolving the position $p$ involves scanning the coded aperture along the z-axis by a certain distance, recording intensities until the intensity modulation pattern uniquely aligns with a pattern within the coded aperture. Due to their property of uniqueness in each sub-sequence of some fixed length, we will employ de Bruijn sequences for this purpose, as we will elaborate on in the section discussing the design of coded apertures. Upon recovering $p$, the second problem of the deconvolution of $s$ can be tackled using a linear solver tailored for the task.

To simulate the intensities in the detector pixel, we approximate the functions $a(z)$ and $s(z)$ using piecewise-constant functions defined on a regular grid. In this discrete setting, the aperture function is represented by the vector $\bm{a} = [a_0, a_2, \ldots, a_{L-1}]^T\in\mathbb{R}^{L}$, where $L$ denotes the grid size, and the coefficients correspond to approximations of optical transmissivity or opacity at corresponding grid intervals. For instance, a value of $1$ signifies unaltered passage of the beam, while a value of $0$ indicates complete signal absorption. In realistic scenarios, these values often lie between $0$ and $1$, and their precise values depend on factors such as the thickness and material composition of the coded aperture, as well as the energy and incidence angle of the incoming beam \cite{Gursoy2022focusing}. Selecting an interval size equal to the step size of the aperture scan can provide a computationally efficient approximation to the measurement process. However, when greater accuracy is required, a smaller interval size can be chosen at the expense of increased computational demands.

The footprint of the beam on the coded-aperture plane, diffracted into the pixel, is approximated by the discrete signal $\bm{s} = [s_0, s_1, \ldots, s_{N-1}]^T\in\mathbb{R}^N$, where $N$ is the discrete size of the diffracted beam. The choice of $N$ is selected based on the signal's extent and the desired depth-resolution. For instance, to accurately resolve a \SI{100}{\micro\meter}-wide signal with an apparent resolution of \SI{1}{\micro\meter}, we need to set $N\geq100$. Additionally, we assume the signal is real-valued because we directly measure intensity, which directly corresponds to the beam's amplitude. In this discrete representation, we can express Equation~\ref{eq1} as either a dot product or a vector product
\begin{equation}
\label{eq1a}
    I = \bm{a}_p  \cdot \bm{s} = \bm{a}_p ^T\bm{s}.
\end{equation}
In essence, $\bm{a}_p = [a_{p}, a_{p+1}, \ldots , a_{p+N-1}]^T\in\mathbb{R}^N$ represents a specific segment of $\bm{a}$ determined by $p$, thereby $N\leq L$. We often utilize identical resolution grids for both the coded aperture and the discrete signal, enabling computationally efficient application of linear operations between them.

We can represent the modulated intensity resulting from scanning the coded-aperture by aggregating intensities at various scan points into a vector, denoted as $\bm{I} = [I_0, I_{1}, \ldots, I_{M-1}]^T\in\mathbb{R}^M$, results from translating the coded-aperture $M$ times. When the translations are equidistant, the measurement formation can be expressed as a matrix product
\begin{equation}
\label{eq2}
\begin{bmatrix}
    I_{0} \\
    I_{1} \\
    \vdots \\
    I_{M-1}
\end{bmatrix}
=
\begin{bmatrix}
    a_{p} & a_{p+1} & \dots  & a_{p+N-1} \\
    a_{p+1} & a_{p+2} & \dots  & a_{p+N} \\
    \vdots & \vdots & \ddots & \vdots \\
    a_{p+M-1} & a_{p+M} & \dots  & a_{p+M+N-2}
\end{bmatrix}
\begin{bmatrix}
    s_{0} \\
    s_{1} \\
    \vdots \\
    s_{N-1}
\end{bmatrix}.
\end{equation}
This relationship can also be expressed compactly as:
\begin{equation}
\label{eq3}
    \bm{I}=\bm{A}_p \bm{s},
\end{equation}
where $\bm{A}_p\in\mathbb{R}^{M\times N}$ is the coding matrix, and $p$ represents the offset scan position, akin to Equation~\ref{eq1a}, signifying the section of the coded-aperture that modulates $\bm{s}$. 

To incorporate experimental constraints like limited exposure time and quantify the robustness of the aperture to noise, we need to model measurement noise. When a photon counting detector is assumed for recording intensities, the recording of intensities follows a Poisson process, which can be expressed as:
\begin{equation}
\label{eq4}
    \bm{d} \sim \operatorname{Poisson}\left\{ \bm{I} \right\},
\end{equation}
where $\text{Poisson}\left\{ \bm{I} \right\}$ denotes an element-wise Poisson distribution with a mean intensity, and the measurement data $\bm{d} = [d_0, d_{1}, \ldots, d_{M-1}]^T$ is a random variable proportional to the incident intensity in the detector pixel. In this scenario, the utilization of photon counting detectors represents the best-case scenario in detection and sets the imaging limits. However, in practical applications, photon integrating detectors are frequently employed despite sacrificing accuracy due to their cost-effectiveness. 

\subsection{Signal recovery model}

In this section, we will outline the procedures for recovering $p$ and $\textbf{s}$ from the measurement data acquired while scanning a known coded aperture. The corresponding reconstruction problem can be formulated as a minimization task given by the equation
\begin{equation}
\label{eq5}
    \min_{p, \bm{s}} \left\| \bm{A}'_p \bm{s} - \bm{d}' \right\|_2^2.
\end{equation}
Here, $\bm{A}'_p$ denotes a normalized coding matrix with coefficients ranging between 0 and 1, as the energy of the diffracted beam remains unknown prior to the experiment. Consequently, we normalize each pixel recording independently, adjusting for relative intensities using the expression, $\bm{d}' = (\bm{d}_{raw}-\mu_0)/(\mu_1-\mu_0)$, where $\mu_0$ and $\mu_1$ are the average intensities corresponding to the 0s and 1s in the raw scan data $\bm{d}$. Because we don't have direct access to the average intensities, we estimate them from the minimum and maximum intensities in the scan data such that $\mu_0 = d_{min} + 2\sqrt{d_{min}}$ and $\mu_1 = d_{max} - 2\sqrt{d_{max}}$, where $2\sqrt{d_{min}}$ or $2\sqrt{d_{max}}$ correspond to two standard deviations from the mean. In scenarios involving unstable illumination sources, additional measurement setups can be employed to monitor and integrate proportional weighting during normalization.

While various methods exist for solving Equation~\ref{eq5}, we adopt a sequential optimization approach that involves updating $p$ while fixing $\bm{s}$, followed by updating $\bm{s}$ while fixing $p$. As the objective function in Equation~\ref{eq5} is non-convex when $\bm{s}$ is fixed, we employ an exhaustive search to find the optimal $p$. This entails evaluating the objective function for all possible signal positions within a feasible range along the coded-aperture, selecting the one that minimizes the objective value. This approach remains robust even when using an approximate representations for $\bm{s}$. For instance, a smoothed boxcar or Gaussian function that approximates the expected grain shape and size can effectively serve as $\bm{s}$ during the optimization process.

For more complex signals, we may employ an alternating optimization strategy to iteratively refine the signal and its position. Calibration of the coded-aperture position before the experiment, particularly when it is close to the sample, can help reduce the computational time for position resolution by reducing the number of potential search positions, although its impact on accuracy is minimal.

Solving for $\bm{s}$ given $p$ essentially constitutes a non-blind deconvolution problem that we can address directly through matrix inversion or, potentially more efficiently with an iterative solver. In this scenario, we construct $\bm{A}'_p$ for the resolved $p$ and solve the linear system of equations to determine $\bm{s}$. When $M$ exceeds $N$ (meaning there is more data than unknown model parameters), the equation set becomes overdetermined, and an exact solution may not exist. Nonetheless, we can obtain an approximate solution in the least-squares sense. Least-squares solutions possess appealing properties, including noise robustness and the availability of widely-used implementations.

In the absence of prior information about the signal, a conservative approach involves collecting as many measurements as there are unknowns to fully describe the signal. Robustness can be ensured by acquiring high signal-to-noise ratio (SNR) measurements through extended exposure times. However, in situations where $M$ is smaller than $N$ or when noise mitigation is necessary, additional constraints, such as non-negativity or enforcing signal smoothness, can be incorporated. These constraints effectively narrow down the solution set by excluding unwanted solutions within the search space. 

\subsection{Design parameters for numerical experiments}

\begin{figure}
\centering
\includegraphics{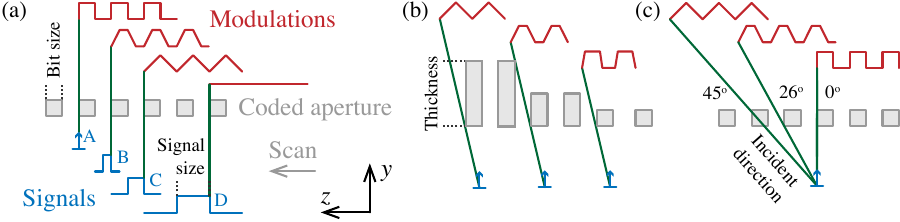}
\caption{Effects of (a) signal size, (b) three aspect ratio or thickness-to-bit size, and (c) beam incident direction on detector pixel intensity modulation (shown in red) with a constant bit size.}
\label{fig2}
\end{figure}

The primary objective behind scanning a coded aperture is to modulate intensity at the detector pixel. Assessing the effectiveness of this encoding involves comprehending the factors influencing this modulation. In Figure~\ref{fig2}, we conceptualize how aperture structural parameters (like bit size and thickness-to-bit ratio) interact with geometry and scanning elements (such as signal size and incident beam direction), impacting the intensity modulation in the detector. Consequently, we focus on structural parameters like the bit-to-signal size ratio, aspect ratio, and incident beam angle to derive conclusions. Additionally, we evaluate individual patterns by correlating the success of reconstruction with specific pattern features, like the mean transmittance within the pattern or the count of bit-flips (or bars) present in the aperture pattern.

Our simulations employed a binary coded-aperture design reminiscent of a barcode, with gold (\ce{Au}) bars used to modulate the diffracted beam. The aperture pattern, derived from a de Bruijn sequence, consisted of 256 bits with an order of 8, ensuring uniqueness within each 8-bit sequence. De Bruijn sequences are particularly well-suited for our purpose due to their property of containing every possible subsequence of a given length exactly once. This characteristic allows for precise control over scanning lengths, making them ideal for meeting the demands of our imaging system.

We selected gold (\ce{Au}) as the material for the aperture due to its ease of workability and desired absorption capabilities within the beam energy range of 5 to \SI{30}{\kilo\electronvolt}. To ensure simplicity and consistency, both 0s (using a thin \ce{Si3N4} membrane with negligible absorption) and 1s (made of \ce{Au}) were assumed to have the same length. The choice of bit sizes and aperture thickness served as control parameters, as we will demonstrate in the subsequent section. This setup was primarily tailored to optimize the operational range for the Laue micro-diffraction microscope at the Advanced Photon Source (APS).

\subsection{Quantification of designs}

Our objective is to accurately reconstruct both the depth and shape of the diffracted beam from measurements. To assess the likelihood of success for a given aperture and scanning setup, we employ a metric, denoted as $S_q$, which compares the recovered parameters ($\bar{\bm{x}}_q$) with their true counterparts ($\bm{x}$), where $q$ represents a specific instance of simulated measurement. This metric, defined as
\begin{equation}
    S_q= 
\begin{dcases}
    1, & \text{if } \left\| \bar{\bm{x}}_q - \bm{x} \right\| < \epsilon \\
    0, & \text{otherwise},
\end{dcases}
\label{eq6}
\end{equation}
quantifies the quality of measurements, classifying instances as successful or not based on their proximity to true values. The error threshold, $\epsilon$, is determined empirically and typically falls within the range of $0.02\pm0.01$, targeting error levels of a few percent.

For the true signal, we selected a bounded Gaussian signal with a size of approximately \SI{10}{\micro\meter}, as this closely resembles signals observed in practical settings and is within the same order of magnitude as the bit sizes of the aperture.

To have a statistical understanding of the overall design, we have to compute $S_q$ for all potential $q$ instances across the aperture and averaging them. Additionally, conducting stochastic trials across different noise instances is needed for testing reliability. Therefore, we introduce the Mean Success Percentage (MSP):
\begin{equation}
    \text{MSP} = 100 \times \frac{1}{K}\sum_{q=1}^{K} S_i 
\end{equation}
where $K$ denotes the total trials. For reliable evaluations, employing 30 trials aligns with the central limit theorem and provides sufficient confidence for assessing the quality of noisy trials, a criterion that we applied in our numerical tests.

\section{Results}

This section offers a comprehensive evaluation of design parameters on the reconstruction performance. We conduct numerical tests to analyze the impact of factors such as bit sizes, scan length, aspect ratio, and coded-aperture patterning. Additionally, we validate our findings through empirical studies using real experimental data. 

\subsection{Selection of the bit sizes}

The size of bits in a coded aperture can significantly affect the total scan length. Smaller bit sizes are favored because they shorten both the scan length and data acquisition time. However, when bit sizes approach or become smaller than the signal size, decoding complexities arise, impacting both position and signal recovery. Hence, our focus remains on bit sizes comparable to the signal to understand the limitations. We introduced the bit-to-signal ratio (BSR), exploring ratios from 0 to 2 across various beam energies and intensity settings. A smaller BSR means that the size of the bit (i.e. the smallest bar in the coded-aperture) is smaller than the extend of the signal incident on the aperture. Specifically, we picked a bounded Gaussian signal size of \SI{10}{\micro\meter} for the true signal and a constant scan length of 8 bits across experiments, resulting in varying scan lengths based on the chosen bit size for a given BSR.

\begin{figure}
\centering
\includegraphics{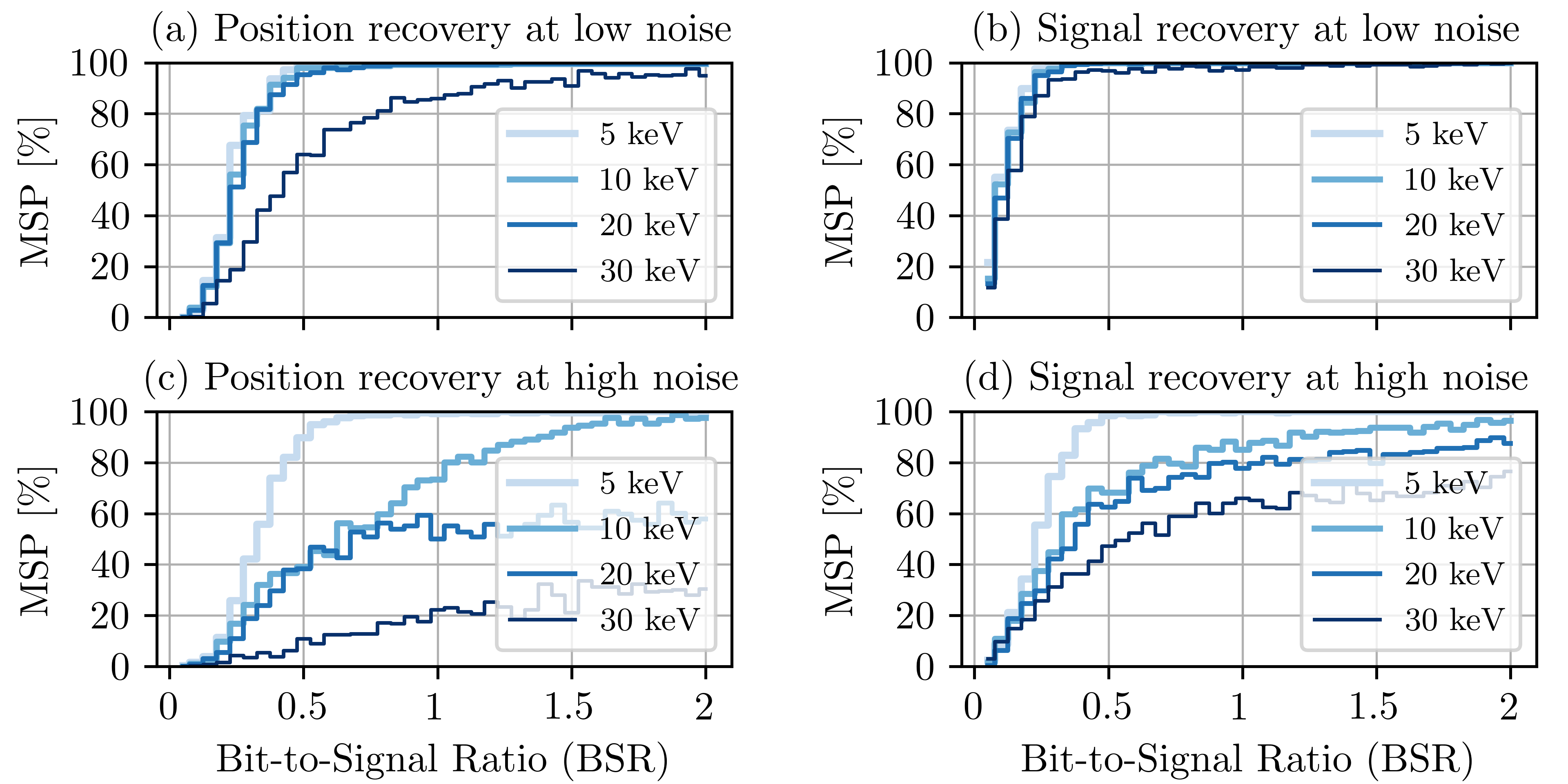}
\caption{MSP plots for position (a and c) and shape (b and d) recovery of the signal at different BSRs. We considered intensity levels: 10 and 100 photon counts per maximum pixel exposure, representing high (c and d) and low (a and b) noise, across varying beam energies.}
\label{fig3}
\end{figure}

Our constant aperture thickness matched the signal size at \SI{10}{\micro\meter}. Employing a fixed step size of \SI{1}{\micro\meter}, larger scans or bit sizes increased the data volume. We simulated acquisitions under two illumination conditions using a photon counting detector and two incident intensity levels: 10 and 100 photon counts per maximum pixel exposure corresponding the high and low noise respectively. We generated scan data from four diffracted beam energies at \SI{5}{\kilo\electronvolt}, \SI{10}{\kilo\electronvolt}, \SI{20}{\kilo\electronvolt} and \SI{30}{\kilo\electronvolt}, typically expected in synchrotron diffraction experiments. We employed the non-negative least squares method through SciPy for solving the position and shape of signal.

Performing simulations 30 times for each potential aperture position, and given that there are 249 unique 8-bit subsequences within a 256-bit sequence, we have a total of $K=30\times249=7470$ instances for computing the Mean Success Percentage (MSP) for position and signal recovery. We used 1-bit size error margin for position recovery and a 0.02 tolerance for signal recovery in calculating the MSP. The resulting MSP plots in Figure~\ref{fig3} demonstrate precise position recovery when BSR exceeds 1, especially in low-noise conditions and energies below \SI{20}{\kilo\electronvolt}. Higher noise levels proportionally reduced MSP with increasing energy. The signal recovery plots mirrored the position recovery trend, showcasing an optimal configuration around $BSR=0.5$, aligning with the idea that smaller bit sizes expedite data collection while precision can be enhanced by appropriately extending the scan length.

\subsection{Scan length}

\begin{figure}
\centering
\includegraphics{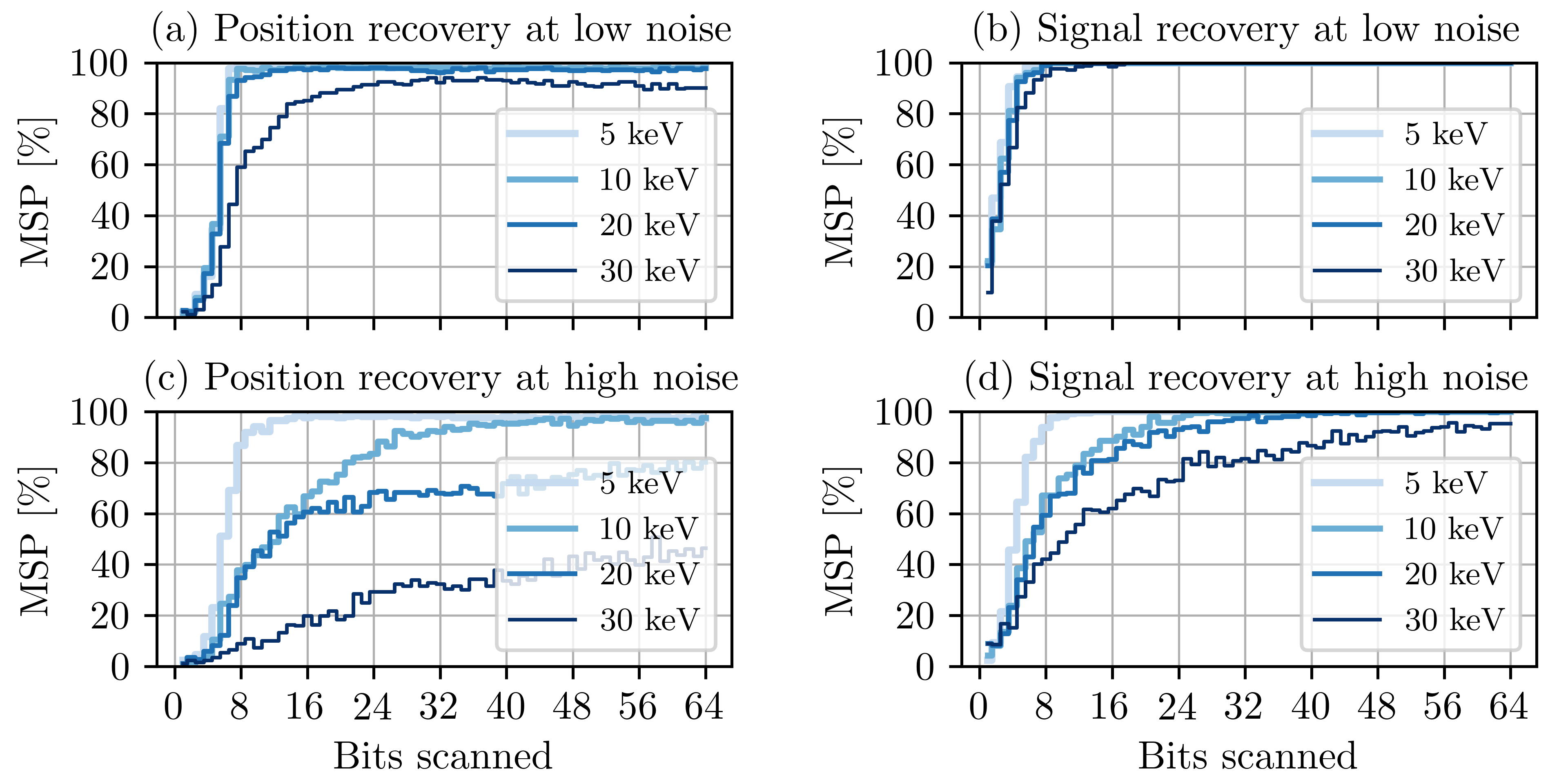}
\caption{MSP plots for position (a and c) and shape (b and d) recovery of the signal at different scanning lengths in terms of bits. We considered intensity levels: 10 and 100 photon counts per maximum pixel exposure, representing high (c and d) and low (a and b) noise, across varying beam energies.}
\label{fig4}
\end{figure}

The plots in Figure~\ref{fig3} provide insights into the relationship between bit size and signal size. It's important to note that in this numerical setting, different bit sizes correspond to varying scan lengths. To examine the impact of increasing scan length on position recovery while keeping the bit size constant, we present the MSP plots in Figure~\ref{fig4}. As anticipated, scanning with fewer than 8 bits leads to significantly lower success rates. Also, higher levels of noise are associated with decreased success rates. Extending the scan length beyond 8 bits does enhance the success rate across all cases, but the rate of improvement is gradual. For instance, in a highly noisy setting, we observe an MSP increase in position recovery from 40\% to 80\% by tripling the scan length for a \SI{10}{\kilo\electronvolt} beam. Yet, it might not sufficiently combat measurement noise unless the scan length is significantly increased. In sum, an optimal scan length for imaging a \SI{10}{\micro\meter}-sized signal appears to be around 8 bits or \SI{80}{\micro\meter}, although a longer scan range might be preferred based on signal size, noise, and experimental factors.

\subsection{Aspect ratio}

\begin{figure}
\centering
\includegraphics{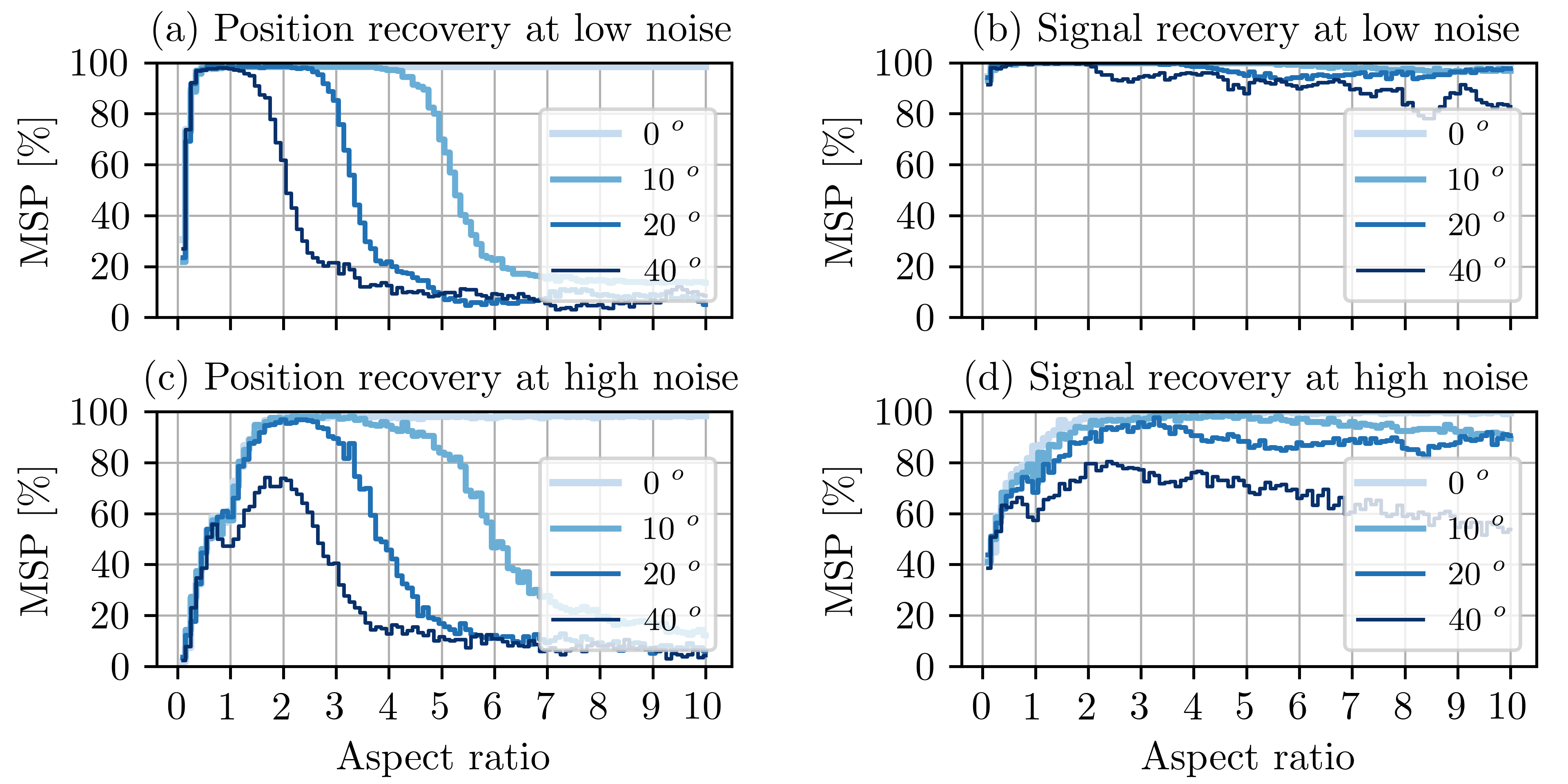}
\caption{MSP plots for position (a and c) and shape (b and d) recovery of the signal at different aperture ratios. We considered intensity levels: 10 and 100 photon counts per maximum pixel exposure, representing high (c and d) and low (a and b) noise, across varying incidence angles.}
\label{fig5}
\end{figure}

Increasing aspect ratio or thickness along the optical axis enhances modulation contrast, especially with a perpendicular incident beam. Therefore, one might consider this as a desirable design choice. However, altering the incidence angle affects the absorption path length of the diffracted beam within the aperture material. Higher angles result in increased absorption likelihood due to a longer traversal distance within the material, causing more complex intensity modulations based on the pattern at these angles.

Figure~\ref{fig5} displays our assessment of aspect ratio for an aperture with a constant bit size of \SI{10}{\micro\meter} on position recovery at four incident angles: \SI{0}{\degree}, \SI{10}{\degree}, \SI{20}{\degree}, and \SI{40}{\degree}, and at varying thicknesses from \SI{1}{\micro\meter} to \SI{100}{\micro\meter}, corresponding to aspect ratios from 0.1 to 10. At \SI{0}{\degree}, representing ideal modulation conditions, MSP peaks except under high noise, where it decreases slightly to 95\%. All plots reach a peak around 0.5 to 2 of aspect ratio. Beyond this optimal point, MSP declines with increasing thickness or aspect ratio. Overall, aspect ratios around 1 generally suit various applications, providing sufficient absorption and enhanced modulation contrast.

\subsection{Patterning of the aperture}

The aperture pattern choice significantly impacts reconstructions. Using our coded aperture, which varies in bar density, we conducted an independent assessment of MSP for each of the 8-bit subsequence in this aperture. In assessment of the MSP results, we used two key statistics in a subsequence: the ratio of 0s to 1s, indicating absorption capability, and bit-flips, quantifying bar variations.

\begin{figure}
\centering
\includegraphics{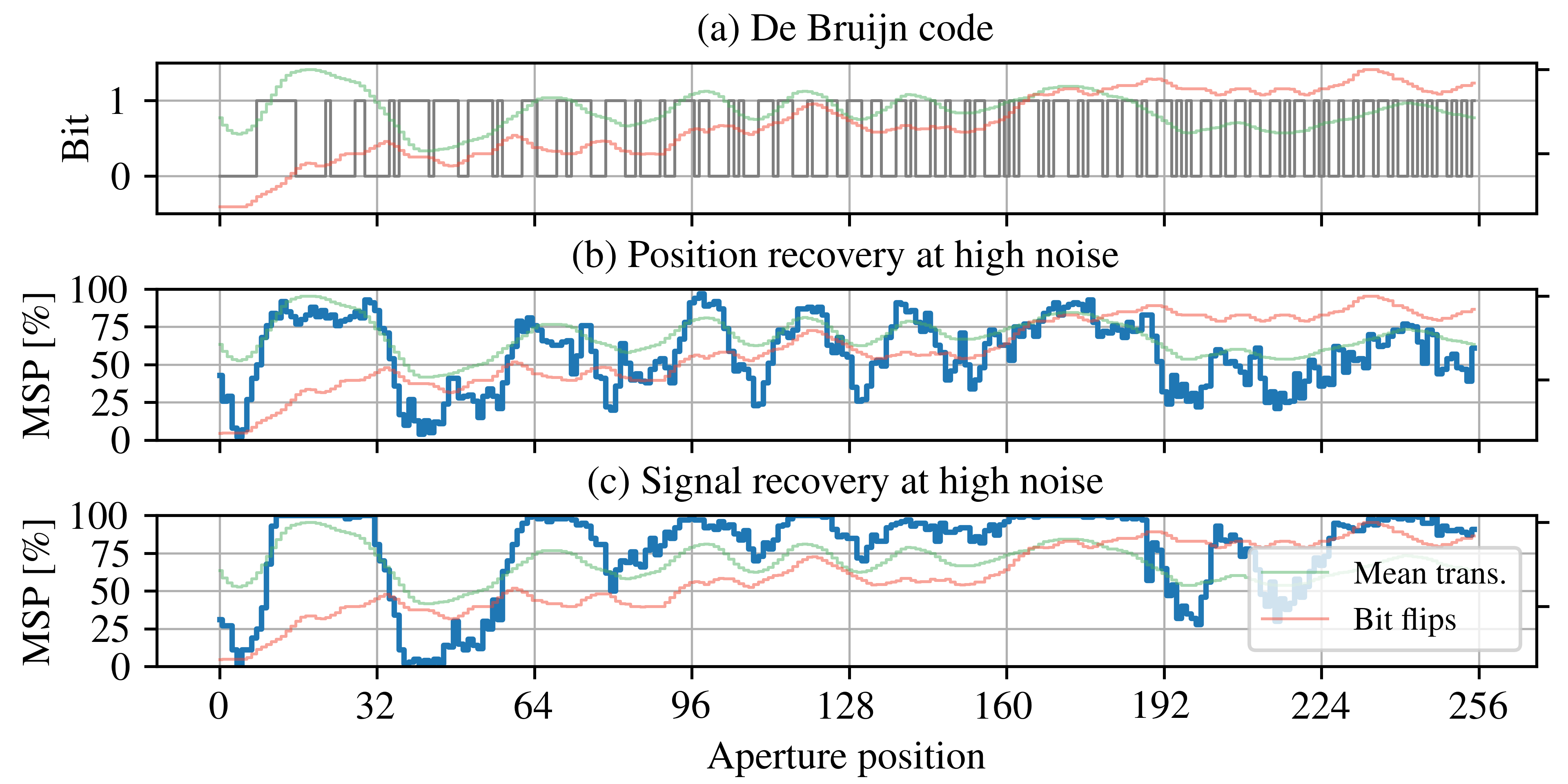}
\caption{(a) De Bruijn code used for signal modulation during scanning. Signal position (b) and shape (c) recovery in high-noise setup. Mean transmittance profile and bit-flips of the 8-bit subsequence are shown in green and red, respectively, in all three plots.}
\label{fig6}
\end{figure}

Figure~\ref{fig6} demonstrates the impact of aperture patterns on position and signal recovery, considering a signal size with a BSR of 0.5. In low noise, recovery remains robust across patterns, minimizing their significance, therefore we haven't showed those plots. In high noise, MSP plots for position and signal recovery highly correlate and are impacted by the ratio of 0s to 1s in the subsequence. More zeros yield more detected photons, benefiting both position and signal recovery. This ratio impacts MSP more than bit flips. In summary, our findings indicate that pattern variation plays a crucial role in photon-starved conditions, with a preference for sequences characterized by a higher frequency of zeros.

\subsection{Empirical studies with experimental data}

\begin{figure}[t]
\centering
\includegraphics{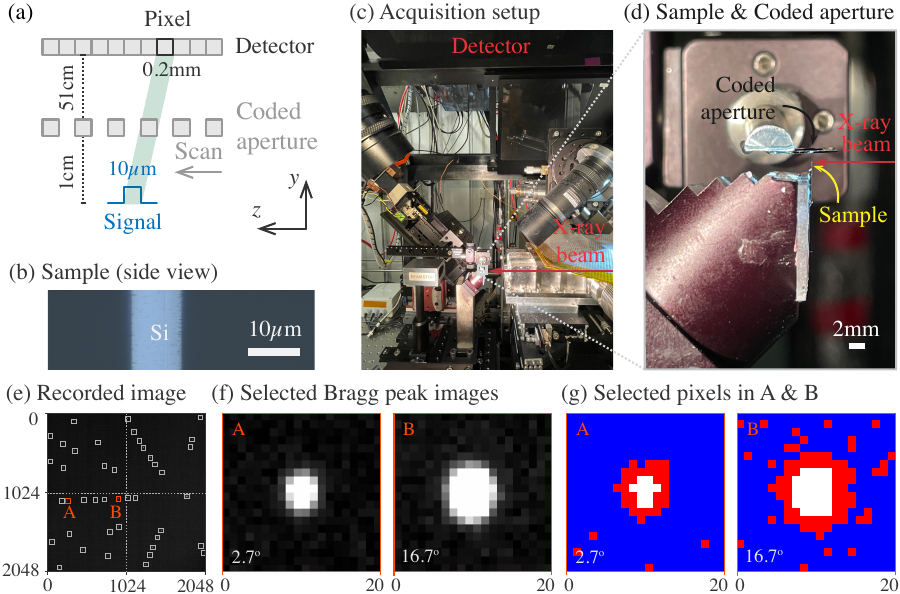}
\caption{(a) Acquisition setup sketch. (b) SEM image of the silicon crystal specimen. (c) Experimental setup with a close-up showing the sample and coded aperture geometry. (d) Laue image captured. (e) Bragg peaks at \SI{2.7}{\degree} and \SI{16.7}{\degree} incidence angles on the coded aperture. (f) Blue and red markers on selected pixels denote intensity levels: 0-20 and 20-200 pixel readout per exposure.}
\label{fig7}
\end{figure}

In our final evaluation, we performed a real experiment at the 34-ID-E beamline at APS, Argonne National Laboratory, employing dedicated Laue diffraction microscopy, as photographed in Figure~\ref{fig7}. This setup utilized non-dispersive Kirkpatrick-Baez mirrors, focusing a polychromatic x-ray beam spanning 7 to \SI{30}{\kilo\electronvolt}, generating a beam spot of approximately \SI{200}{\nano\meter} in diameter on the sample. We used a coded aperture identical to that in Figure~\ref{fig6}, fabricated at the Center for Nanoscale Materials, Argonne National Laboratory, through direct-write lithography. Comprising roughly \SI{4.6}{\micro\meter} thick gold bars on a thin silicon nitride (\ce{Si3Ni4}) membrane, the aperture had bit sizes of about \SI{15}{\micro\meter} and \SI{7.5}{\micro\meter} for 0s and 1s, respectively. 

To obtain the true signal, we utilized diffraction from a silicon single crystal, which was \SI{10}{\micro\meter} thick and oriented perpendicular to the incident x-ray beam (refer to Figure~\ref{fig7}). The coded aperture was placed approximately \SI{0.84}{\milli\meter} above the sample and systematically raster scanned with a step size of \SI{1}{\micro\meter}, covering a distance of \SI{600}{\micro\meter}. Laue images were captured using a Perkin-Elmer detector with dimensions of $409.6\times\SI{409.6}{\milli\meter}^2$, featuring $2048\times2048$ pixels and a 16-bit dynamic range. Configured in a \SI{90}{\degree} reflection geometry, the detector was situated \SI{510.9}{\milli\meter} above the sample to capture the Laue patterns. It's important to note that the detector operates on a photon-integrating principle, and thus, the measurement has contamination from detector noise.

In our measured Laue diffraction image, we picked two square regions of interest with 20$\times$20 pixel size capturing Bragg peaks on the detector, as marked in Figure\ref{fig7}. Those Bragg spots corresponds to incidence angles of \SI{2.7}{\degree} and \SI{16.7}{\degree}. The pixels in these regions were sorted into two bins: 0-20 and 20-200 values, representing high and low noise, respectively. We computed MSP for these bins to evaluate position and signal recovery accuracy across scan lengths from 10 to \SI{600}{\micro\meter} (approximately 40 bits). 

\begin{figure}
\centering
\includegraphics{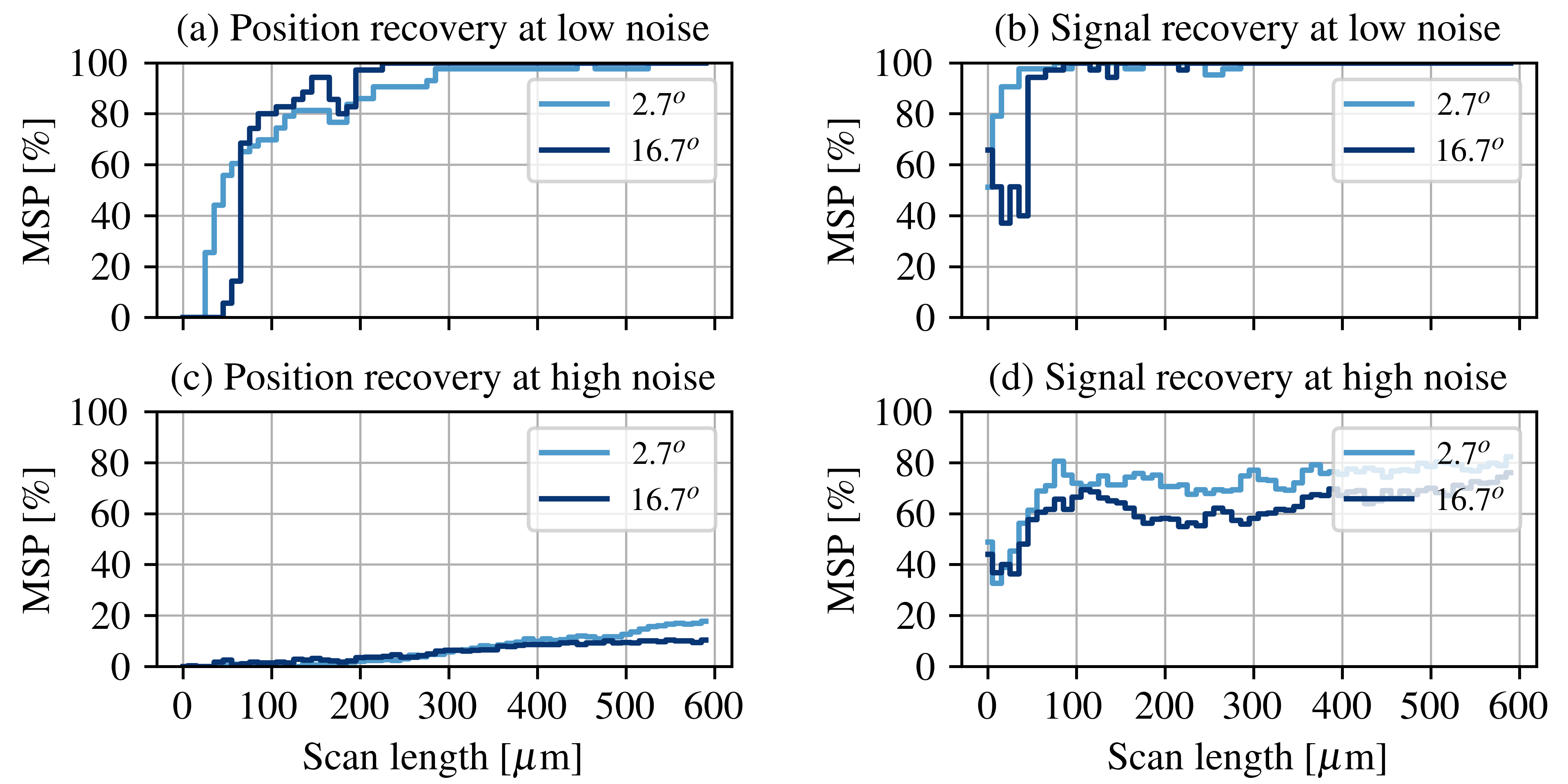}
\caption{MSP plots for position (a and c) and shape (b and d) recovery of the signal at different scanning lengths. 8 bit range is marked with dotted line. We considered intensity levels: 0-20 and 20-200 pixel readout per exposure, representing high (c and d) and low (a and b) noise at two incidence angles of of \SI{2.7}{\degree} and \SI{16.7}{\degree}.}
\label{fig8}
\end{figure}

Figure~\ref{fig8} shows the corresponding MSP plots. Considering our present bit sizes of 7.5 and \SI{15}{\micro\meter}, achieving a unique encoding requires a \SI{120}{\micro\meter} scan range for an 8-bit sequence. In low noise conditions, we attained high accuracy for both position and signal recovery. Extending beyond 8 bits (\SI{120}{\micro\meter}), disparities between the two angles had negligible impact. However, in high noise, MSP for position recovery was lower than predicted, likely due to the inclusion of pixels with readouts below 10. Furthermore, our scintillator-based detector introduced background noise, and the consecutive readout values of pixels were correlated due to a non-zero scintillation decay time, impacting performance in high-noise scenarios. The difference between the incidence angles were as expected; an angle of \SI{2.7}{\degree} yielded a higher MSP compared to \SI{16.7}{\degree} overall. With longer scan lengths, up to 40 bits, we observed a slight improvement in quality.

In conclusion, we found that an 8-bit minimum scan length delivers adequate MSP for diverse applications, and the incidence angle poses fewer issues for aspect ratios below 1 in the default Laue microscope setup.

\section{Discussions}

The success of our experiments depends on numerous factors, extending beyond just the aperture design to encompass scanning methodology and subsequent analytical techniques. Consequently, considering all these factors comprehensively in experimental design, which includes apertures, scanning protocols, and algorithms, poses a significant challenge, particularly when conducted through numerical simulations. Achieving a more exhaustive optimization of these parameters necessitates the utilization of high-performance computing capabilities. This allows for the exploration of optimal designs through large-scale computations. Given that the signal reconstruction problem is independent for each detector pixel, leveraging parallel and distributed computing approaches \cite{prince2023cross} emerges as a feasible strategy.

Initially, we chose equally spaced data sampling along the scan direction of the coded aperture. However, an alternative approach, not currently in practice, involves utilizing non-equidistant sampling along the scan length. This method shows promise when coupled with "smart" sampling techniques, which can become a crucial element within a reinforcement learning strategy. By integrating this approach, dynamic adjustments in the sampling process become possible, allowing for optimization of data collection based on feedback and learned patterns. This, in turn, could lead to improved efficiency and accuracy in the overall scanning process.

We utilize a standard Euclidean norm to define a cost function, which follows a least squares approach, serving as the foundation of our simulations. However, alternative methods could potentially offer more robust solutions. For example, incorporating stochastic models that compute likelihood functions to represent the formation of data could lead to a more adaptable construction of the cost function. There are few computationally feasible options available, such as adopting Gaussian or Poisson assumptions for the detected intensities, each with its own set of advantages. The Poisson model tends to produce superior results, particularly when employing photon counting detectors near their detection limit, while the Gaussian approximation provides simplicity in optimization and can enhance overall accuracy, especially for photon integrating detectors.

In addition to considering likelihood image formation models, we can also integrate prior information about the signals of interest to further refine our solutions \cite{Gursoy:15}. This process, known as "regularization" or "Bayesian inference" in certain contexts, involves leveraging existing knowledge about signal characteristics or imposing constraints on their properties. These constraints may include generic properties like non-negativity and expected boundaries, or leveraging learning models such as dictionary learning or deep neural networks for the task. By incorporating this prior information into the reconstruction process, we refine and narrow down potential solutions, thereby enhancing the accuracy and robustness of the final outcomes.

The process of signal recovery occurs sequentially. Initially, we determine the signal's direction by matching the modulation pattern on the coded aperture with the signal's position, then deduce the signal's shape at its source. Consequently, the effectiveness of signal recovery hinges on accurately pinpointing the position. To enhance the precision of position detection, calibrating the coded-aperture's positioning before the experiment becomes important. This calibration restricts the feasible locations for the beams to pass from the sample to the detector pixel, thereby narrowing down the range of possible subsequences in the code that require examination. This strategy holds significant potential for enhancing position accuracy, particularly in high-noise settings, thereby improving the overall signal recovery process.

In this study, we deliberately focused exclusively on binary coded-apertures while excluding non-binary alternatives. The rationale behind this choice is rooted in the fabrication process, where non-binary coded-apertures introduce additional complexities and are susceptible to errors that may result in structural artifacts within the aperture. Our intention is to first thoroughly investigate the physical errors and their origins in the aperture fabrication process, and subsequently develop appropriate models to address them before delving into the assessment of non-binary apertures or making comparisons with their binary counterparts. It's worth noting, however, that once an accurate model and scanning configuration are established, the methods detailed in this paper can be readily applied to evaluate and analyze any type of aperture design.

Expanding the scope of micro-scale coded apertures prompts exploration into multi-thickness and multi-material designs, opening avenues for enhanced functionalities. Multi-thickness apertures exhibit variable depths within individual elements, enabling improved control over light modulation and encoding complexity. Similarly, integrating diverse materials can offer unique optical properties, enabling tailored functionalities like polarization filtering or wavelength selectivity. However, these innovations introduce complexities in fabrication, demanding intricate processes capable of accurately patterning different materials and thicknesses.

Various fabrication techniques cater to micro-scale coded aperture production, each with distinct advantages and limitations. Photolithography, commonly used in semiconductor manufacturing, offers high precision but faces challenges in achieving extreme aspect ratios and multi-material patterning. On the other hand, focused ion beam (FIB) milling allows for intricate designs with higher aspect ratios but suffers from slower fabrication speeds and material redeposition issues. Additive manufacturing techniques like 3D printing show promise in creating complex, multi-material structures but might lack the resolution required for high-fidelity micro-scale apertures.

Addressing these challenges requires a delicate balance between targeted resolution, fabrication complexity, and material constraints. Advancements in fabrication techniques and materials offer promising solutions to overcome these limitations, paving the way for the creation of more effective and functional coded aperture designs at micro scales. Additionally, precision engineering efforts in instrumentation can improve control and accuracy in positioning during scanning, ultimately resulting in higher-resolution imaging. Furthermore, developments in signal processing algorithms tailored for specific sample types and data acquisition protocols can facilitate robust reconstructions and expedite data collection processes.

\section{Conclusions}

In summary, our study provided insights into the design choices of depth-resolved diffraction experiments with coded-apertures. Typically, bit sizes comparable to or larger than the signals, aspect ratios near 1, and an 8-bit minimum scan suffice for most applications with a decent signal-to-noise ratio. For experiments with limited photons, extending the scan range and employing photon counting detectors could enhance accuracy in capturing modulated intensities. We believe our findings offer a valuable resource for advancing high-speed diffraction experiments and inspiring tailored experimental designs for specific applications.

\section*{Acknowledgements} This research used resources of the Advanced Photon Source and the Center for Nanoscale Materials, U.S. Department of Energy (DOE) Office of Science user facilities at Argonne National Laboratory and is based on research supported by the U.S. DOE Office of Science-Basic Energy Sciences, under Contract No. DE-AC02-06CH11357.

\bibliographystyle{ieeetr}
\bibliography{refs}

\end{document}